**Title:**

Radio-frequency detected fast charge sensing in undoped silicon quantum dots

**Authors:**

Akito Noiri[1*†], Kenta Takeda[1†], Jun Yoneda[1#], Takashi Nakajima[1], Tetsuo Kodera[2] and Seigo Tarucha[1*]

**Affiliations:**

[1]*RIKEN, Center for Emergent Matter Science (CEMS), Wako-shi, Saitama 351-0198, Japan*

[2]*Department of Electrical and Electronic Engineering, Tokyo Institute of Technology, O-okayama, Meguro-ku, Tokyo 152-8552, Japan*

[#]Present address: *School of Electrical Engineering and Telecommunications, University of New South Wales, Sydney, New South Wales 2052, Australia*

[†]These authors contributed equally to this work.

[*]e-mail: akito.noiri@riken.jp or tarucha@riken.jp

**Abstract:**

Spin qubits in silicon quantum dots offer a promising platform for a quantum computer as they have a long coherence time and scalability. The charge sensing technique plays an essential role in reading out the spin qubit as well as tuning the device parameters and therefore its performance in terms of measurement bandwidth and sensitivity is an important factor in spin qubit experiments. Here we demonstrate fast and sensitive charge sensing by a radio-frequency reflectometry of an undoped, accumulation-mode Si/SiGe double quantum dot. We show that the large parasitic capacitance in typical accumulation-mode gate geometries impedes reflectometry measurements. We present a gate geometry that significantly reduces the parasitic capacitance and enables fast single-shot readout. The technique allows us to distinguish between the singly- and doubly-occupied two-electron states under the Pauli spin blockade condition in an integration time of 0.8 μs, the shortest value ever reported in silicon, by the signal-to-noise ratio of 6. These results provide a guideline for designing silicon spin qubit devices suitable for the fast and high-fidelity readout.

Electron spins confined in semiconductor quantum dots (QDs) are an attractive platform for quantum information processing [1]. In particular, silicon (Si) QD devices have been employed to realize single- and two-qubit entangling gates with high-fidelity by virtue of their long coherence time [2–7]. However, high-fidelity qubit readout remains challenging as it requires fast readout compared to the spin relaxation time. Moreover, in order to perform measurement-based protocols such as error correcting codes [8–10], qubit readout within a dephasing time of 1 to 100 μs [2–7] is of particular importance. A qubit readout is commonly performed using a charge sensing technique [11,12] in conjunction with a spin-to-charge conversion process such as spin-dependent electron tunneling between a QD and the adjacent reservoir [13,14] and between QDs in Pauli spin blockade (PSB) [15–17]. Especially for the latter scheme (singlet-triplet readout), the spin-to-charge conversion can be fast ($< 1$ μs) and accurate (conversion fidelity $> 99$ %) [18]. Then the measurement fidelity and bandwidth of the charge sensing rather restrict the performance of the qubit readout. Furthermore, the charge sensing technique plays an important role in optimizing device parameters such as charge occupancy and tunnel couplings. After all, it is highly desired in spin qubit experiments to develop fast and reliable charge sensing with the time resolution $< 1$ μs and the charge discrimination fidelity $> 99$ % (corresponding to signal-to-noise ratio (SNR) $> 5.2$).

The charge sensing signal is commonly obtained by measuring the source-drain current of a capacitively-coupled sensor transistor implemented either by a quantum point contact or an additional QD fabricated near the detected QD [4–7]. The measurement bandwidth of the sensor current usually has a high-frequency cutoff of typically less than 100 kHz caused by the RC network of cryogenic wiring and the IV (current-to-voltage) converter. Use of a cryogenic IV converter can improve the bandwidth up to 1 MHz [19] but a rather long integration time of 12.5 μs is necessary to achieve a high-fidelity singlet-triplet readout (SNR of 6.5) [20,21]. To improve the performance of charge sensing, a radio-frequency (rf) reflectometry technique was developed [22,23]. Here the rf tank circuit is connected to the sensor contact to detect the sensor conductance change. This technique was initially implemented for standard depletion-mode GaAs QD devices to detect the charge states in single and multiple QDs [22–24]. Indeed, inter-dot charge transitions are well resolved in a single-shot manner with a SNR exceeding 6 in an integration time of 1 μs [23]. However, the reflectometry technique is not directly applicable to undoped, accumulation-mode QDs fabricated in standard geometries [16,25], widely used for high-fidelity Si spin qubit devices. Those devices tend to have a parasitic capacitance of several pF between the accumulation gate and the gate-induced two-dimensional electron gas (2DEG) in a Si quantum well [26,27], which pushes the tank circuit in a typical operation condition of the charge sensor away from the impedance matching with the 50 Ohm coaxial feed line. One solution to this problem is to connect the rf tank circuit to the accumulation gate instead of the sensor source contact [26]. This technique is applicable to undoped Si spin qubit devices without having a constraint on device designs. However, the demonstrated sensitivity remains insufficient to realize single-shot spin readout within 1 μs. Note that another type of charge sensor which directly detects quantum capacitance change of a QD associated with charge reconfigurations has recently been developed [28–31]. This technique may be advantageous in terms of scalability because it does not require an external charge sensor, but its sensitivity tends to be worse than that of conductance measurement.

In this Letter, we realize fast and sensitive rf reflectometry measurements of Si/SiGe QDs. This is achieved by introducing a specially designed device geometry comprising a small accumulation gate area ($\sim 10^2$ μm$^2$) to reduce the parasitic capacitance [32]. With this device geometry, we can obtain a good impedance matching in a condition appropriate for charge sensing, and therefore the reflectometry signal is highly sensitive to charge configurations. We demonstrate that singly- and doubly-occupied two-electron states under the PSB condition can be distinguished by a SNR of 6 in an integration time of 0.8 μs. This technique is also useful for other accumulation-mode devices including Si-MOS QDs.

To investigate a device geometry suitable for the rf-detected charge sensing, we design and fabricate two types of undoped, accumulation-mode Si/SiGe double QD (DQD) devices with different accumulation gate geometries (Figures 1(a) and (b)). The devices are fabricated on a commercially grown, isotopically natural Si/Si$_{0.7}$Ge$_{0.3}$ heterostructure as shown in Figure 1(c). The accumulation gate is used to induce a 2DEG in the Si quantum well, while the fine gates are used to control the confinement to form the DQD and the sensor QD. The Ohmic contacts are fabricated by phosphorus (P) ion implantation followed by an activation annealing. The devices are mounted on a printed circuit board and one of the sensor Ohmic contacts (O3) is connected to an LC resonant circuit consisting of a chip inductor ($L = 1.2$ μH) and a parasitic capacitance $C_\mathrm{p}$ as shown in Figure 1(d). The reflection of the rf carrier at frequency $f_\mathrm{rf}$ depends on the matching between the complex impedance of the resonant circuit $Z(f_\mathrm{rf}) = i2\pi f_\mathrm{rf} L + 1/(G_\mathrm{CS} + i2\pi f_\mathrm{rf} C_\mathrm{p})$ and the 50 Ohm characteristic impedance $Z_0$ of the coaxial cable (Figure 1(e)). Here $G_\mathrm{CS}$ is the conductance of the charge sensor. Since the reflected signal is sensitive to $G_\mathrm{CS}$ only around the matching conductance of $G_\mathrm{match} \sim Z_0 C_\mathrm{p}/L$, $G_\mathrm{match}$ has to be comparable to the typical working point of the charge sensor (less than a conductance quantum $e^2/h$). With the realistic range of $L$ (up to a few μH), this restricts $C_\mathrm{p}$ up to $\sim 1$ pF for a sensitive charge sensing by the rf reflectometry.

The accumulation gate of device 1 covers a large area ($\sim 10^4$ μm$^2$) to connect the induced 2DEG to Ohmic contacts (Figure 1(a)), as typically does in accumulation-mode Si QD devices [25,33]. The accumulation gate has a capacitive coupling to the sensor source Ohmic contact in the overlapping region (highlighted in red in Figure 1(a)), estimated to be 1.3 pF by design. In addition, when the 2DEG is accumulated, a capacitance as large as 3.8 pF is induced between the 2DEG and the gate (highlighted in green in Figure 1(a)). These capacitances contribute to $C_p$, preventing the rf reflectometry measurement. To suppress these capacitances, we fabricate devices having a smaller accumulation gate (device 2) by extending the ion-implanted region as shown in Figure 1(b). As a result, $C_p$ due to the accumulation gate is reduced down to 0.02 pF by design, enabling sensitive rf reflectometry charge sensing.

We first characterize $f_{rf}$ dependence of the reflectometry signal in the large (device 1) and small (device 2) accumulation gate devices. The measurement is performed in a 2 K insert. A homemade cryogenic amplifier with a gain of ∼ 40 dB is placed at the base temperature. Note that the temperature increases up to 3.3 K when the amplifier is on. An rf carrier of -10 dBm from a vector network analyzer (VNA) is attenuated by 30 dB at room temperature and further attenuated by 40 dB inside the insert before being applied to the device through a directional coupler with a coupling of -15 dB. The reflected carrier is amplified by the cryogenic amplifier followed by 3 dB attenuation at room temperature and finally measured by the VNA.

Figure 2(a) shows the reflection $S_{21}(f_{rf})$ of the rf signal with and without the accumulated 2DEG in device 1. Below the 2DEG accumulation threshold, the resonance dip is observed at $f_{res} = 125.2$ MHz, yielding $C_p = 1.3$ pF. Above the accumulation threshold, the resonance dip shifts to $f_{res} \sim 72$ MHz, yielding $C_p \sim 4.0$ pF. The obtained values of $C_p$ mostly agree with those expected for the accumulation gate geometry. Slight deviation of the capacitance values could be due to the dielectric constant of our $Al_2O_3$ insulating layer being lower than expected value of 9. Then we measure a pinch off curve of the charge sensor as a function of the sensor plunger gate voltage $V_{SP}$ as shown in Figure 2(b). Figure 2(c) shows $S_{21}(f_{rf})$ measured in the same range of $V_{SP}$, showing no discernible dependence on the sensor conductance due to the impedance mismatch. These results indicate the geometry of device 1 is incompatible with the rf reflectometry measurement.

The resonance frequency of device 2, in contrast, shifts only by 0.6 MHz when a 2DEG is accumulated under the gate (Figure 2(d)). The obtained $f_{res} = 197.5$ MHz and $f_{res} = 196.1$ MHz correspond to $C_p = 0.54$ pF and 0.55 pF respectively, in agreement with the design. These values are much smaller than those in device 1, and are mostly due to the off-chip components such as bonding wires and the parasitic capacitance of the chip inductor. Similar to Figure 2(b), we measure the pinch off curve of the charge sensor in Figure 2(e) and its reflectometry response in Figure 2(f). We observe a clear change of $S_{21}$ by modifying $G_{CS}$ at the resonance condition of $f_{res} = 196.1$ MHz (Figures 2(g) and 2(h)). The matching condition is satisfied at $V_{SP} = 0.55$ V and $G_{CS} \sim$ 0.2 $e^2/h$, where $S_{21}$ drops by more than 30 dB from a maximum and changes sensitively with $G_{CS}$ (Figure 2(h)). This enables a sensitive charge sensing by rf reflectometry measurement. From these results, we confirm the geometry of device 2 is compatible with the rf reflectometry measurement. In addition, we verify that the device geometry dose not degrade the performance of spin qubit manipulation or the controllability and stability of the DQD (see Supporting Information Section 2).

To investigate the performance of the single-shot charge readout using the rf reflectometry, we measure another device whose structure is nominally identical to that of device 2 in a dilution refrigerator with an electron temperature of 120 mK. Here we use $L = 1.0$ μH and obtain $f_{res} = 206.7$ MHz, indicating $C_p = 0.6$ pF. The rf carrier from a signal generator is attenuated in the external circuit both in room temperature environment and inside the refrigerator to achieve an optimal power at the sample end ($\sim -100$ dBm). The reflected signal is first amplified by a cryogenic amplifier (Weinreb CITLF1) with a gain of $\sim 45$ dB placed at the 4 K stage and further amplified at room temperature followed by demodulation using a mixer. Then the mixer output is amplified with an IF amplifier and the signal $V_{rf}$ is recorded by a digitizer.

We form a few electron DQD and observe the charge states in the PSB regime using the reflectometry setup. Figure 3(a) shows a stability diagram measured with a three-step voltage pulse (Figure 3(b)) applied continuously [34]. At pulse stage R, we randomly initialize the two-electron state to one of the four spin states, a singlet and three triplet states. While the triplet states stay in (1,1) charge state at stage M due to the PSB, the singlet state transitions to (0,2). Here (*n,m*) represents the numbers of electrons inside the left (*n*) and right (*m*) QD. The PSB signal is observed inside the trapezoidal region in the (0,2) charge state (Figure 3(a)). Then we measure the PSB signal in a single-shot manner. Figure 3(c) shows a histogram of $V_{rf}$ integrated for time $t_{int} = 0.8$ μs at stage M. We observe clear two peaks corresponding to (0,2) and (1,1) charge states, which enables fast singlet-triplet readout. Fitting with two Gaussian distributions yields the SNR of 6.0. The SNR increases with $t_{int}$ (Figure 3(d)) and the charge discrimination fidelity reaches 99.99 % at $t_{int} = 1.8$ μs. This performance is considerably better than those of the previous experiments in Si [20,21,26,29–31] and comparable to the highest sensitivity ever reported for GaAs devices [23].

In conclusion, we demonstrate how to perform a fast charge sensing in an accumulation-mode Si/SiGe QD devices by embedding a charge sensor in an rf tank circuit. We compare the availability of rf reflectometry technique between two devices in different accumulation gate geometries. The device in a conventional design shows a large parasitic capacitance due to the accumulation of the 2DEG. We find this capacitance leads to the impedance mismatch and impedes the charge sensing by the rf reflectometry. In contrast, this capacitance is suppressed in the device having a small accumulation gate, keeping the matching condition close to the working point of the charge sensor. The reflectometry technique allows us to perform singlet-triplet readout in a single-shot manner with a SNR of 6 in an integration time of 0.8 μs which is much faster than a typical relaxation time (> 10 ms) [4,5] of Si spin qubits at the condition away from spin-valley mixing [35,36] and will be useful for implementing measurement-based protocols. Our technique will be applicable to a variety of accumulation-mode devices including Si MOS QDs, allowing for fast and high-fidelity readout of spin qubits in conjunction with existing techniques of efficient spin-to-charge conversion [17,18,37]. With the demonstrated high-fidelity qubit control [2–7], our device design will further push the high performance of Si-based quantum computers.

**ACKNOWLEDGEMENT**

We thank the Microwave Research Group in Caltech for technical support. This work was financially supported by Core Research for Evolutional Science and Technology (CREST), Japan Science and Technology Agency (JST) (JPMJCR15N2 and JPMJCR1675), and MEXT Quantum Leap Flagship Program (MEXT Q-LEAP) (JPMXS0118069228). A.N. acknowledges support from JSPS KAKENHI grant number 19K14640. K.T. acknowledges support from JSPS KAKENHI grant number JP17K14078. T.N. acknowledges support from JSPS KAKENHI grant number 18H01819 and The Murata Science Foundation. S.T. acknowledges support from JSPS KAKENHI grant numbers JP26220710 and JP16H02204.

**Figures:**

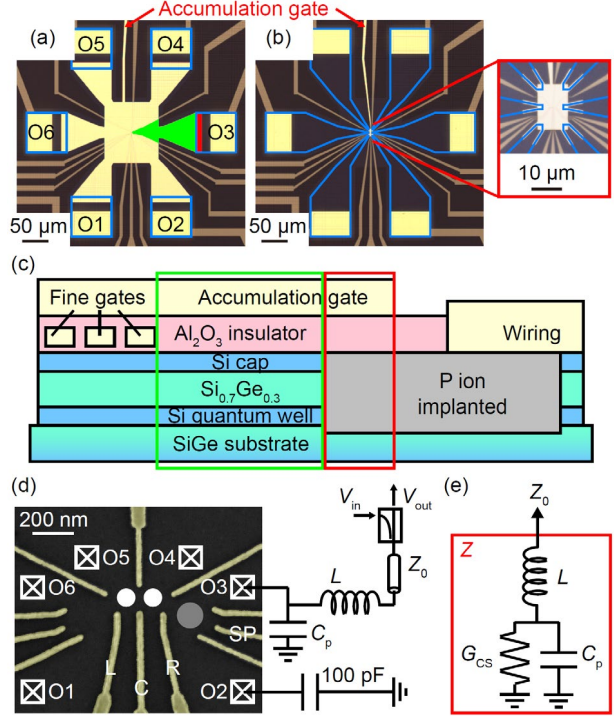

**Figure 1.** Device structure and rf reflectometry setup. (a) Optical microscope image of a device having a large accumulation gate. P ion-implanted regions are marked by blue squares. Green and red regions [their layer stacks are shown in the colored squares in (c)] have large parasitic capacitances contributing to $C_\text{p}$. (b) Optical microscope image of a small accumulation gate device. The areas surrounded by the blue traces show P ion-implanted regions. Inset: Zoom-in image where the small accumulation gate overlaps the implanted regions. (c) Schematic of the device layer structure. The 2DEG is formed at the Si quantum well which is contacted to the P ion-implanted region (Ohmic contact). All of the metallic gates and connection wirings are made of Ti (10 nm thick) and Au (20, 40 and 150 nm thick for the fine gates, the accumulation gate and connection wirings, respectively). The thickness of $Al_2O_3$ insulating layer grown by atomic layer deposition, Si cap, $Si_{0.7}Ge_{0.3}$ spacer, and Si quantum well is 60 nm, 2 nm, 60 nm, and 15 nm, respectively. Capacitance values due to the accumulation gate (see main text) are estimated assuming plate capacitors of these layer thicknesses with the dielectric constant of 9 for $Al_2O_3$, 12 for Si, and 13 for $Si_{0.7}Ge_{0.3}$, respectively. (d) False color scanning electron microscope image of a device. An rf tank circuit is connected to the upper Ohmic (O3) of the charge sensor. (e) Equivalent circuit model of (d).

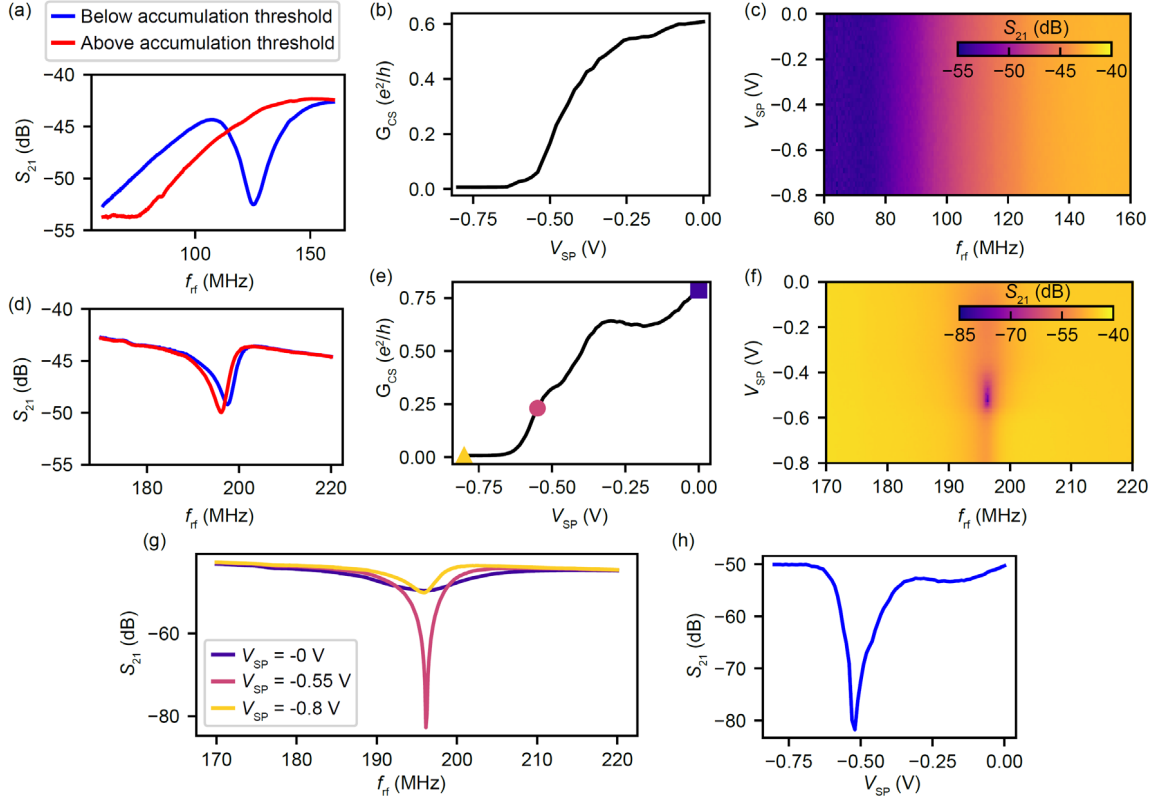

**Figure 2.** Reflectometry response measured with devices in different accumulation gate geometries. (a) Frequency $f_{rf}$ dependence of the reflection $S_{21}$ measured at the accumulation gate voltage below (blue) and above (red) the accumulation threshold for device 1. $S_{21}$ reduction at $f_{rf} < 100$ MHz is due to the gain property of the cryogenic amplifier. (b) Dependence of the charge sensor conductance on the sensor plunger gate voltage $V_{SP}$ for device 1. The sensor conductance $G_{CS}$ is calculated by measuring the sensor current with a d.c. voltage bias of 200 μV, which is applied to O3 through a bias tee on the sample board while the other Ohmic contacts are grounded. (c) $S_{21}$ as a function of $V_{SP}$ and $f_{rf}$ for device 1. (d)-(f) Same plots as in (a)-(c) measured for device 2. (g) Line cuts of (f) with the sensor condition open (purple), matching (magenta) and pinch off (yellow), respectively. They are measured at the colored marker positions in (e). (h) Trace of (f) at the resonance condition of $f_{rf} = 196.1$ MHz.

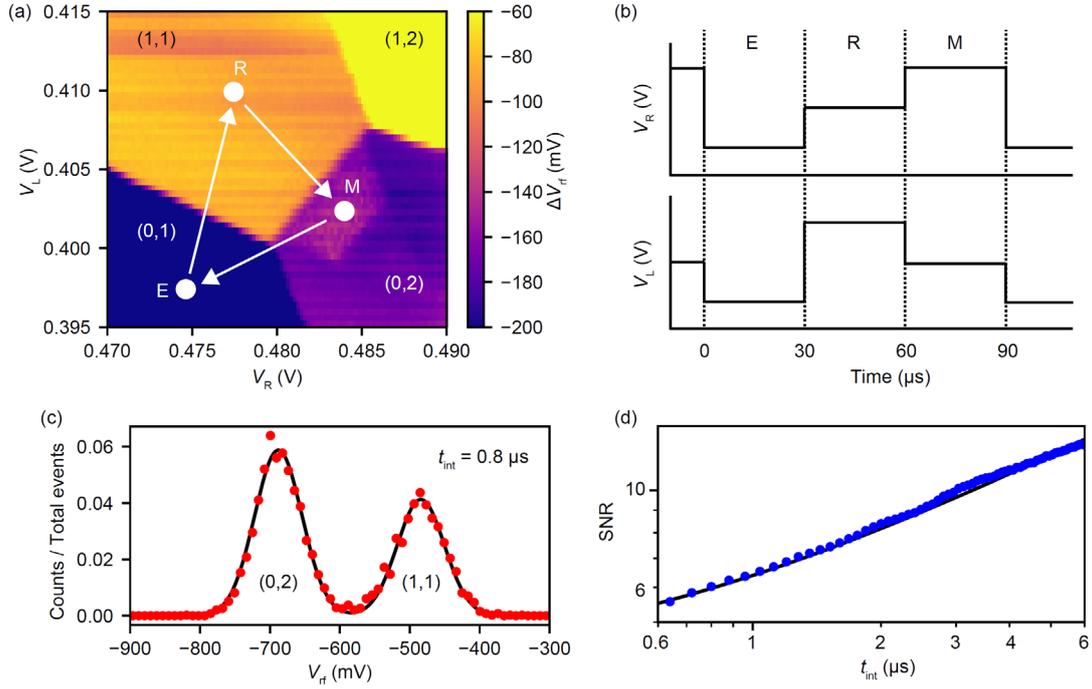

**Figure 3.** Performance of rf-detected charge sensing. (a) Rf-detected stability diagram of a few electron DQD as a function of the gate voltages $V_\text{L}$ and $V_\text{R}$ applied to the gate electrodes L and R, respectively. An in-plane external magnetic field of 0.2 T is applied. A three-step pulse to change the gate voltage conditions for evacuation (E), initialization (R) and measurement (M) is applied continuously [34]. Except for the measurement stage (M), the rf carrier is blanked. A plane is subtracted from the raw data to remove the sensor signal change by the gate voltages. (b) Schematic of the voltage pulse shape. (c) Histogram of single-shot measurements at an integration time of $t_\text{int} = 0.8$ μs and the external magnetic field of 0.6 T. The pulse dwell times at individual points are changed to 50 μs (E), 5 μs (R) and 10 μs (M), respectively. The black curve is a fit with a sum of two Gaussian distributions. From the fit we obtain a SNR of 6.0. Here the signal is the difference between the two peaks and the noise is the standard deviation of the Gaussians. (d) $t_\text{int}$ dependence of the SNR. The black curve is a fit with $\text{SNR}(t_\text{int}) = a \times \sqrt{t_0 + t_\text{int}}$ with $a = 5.1$ μs$^{-1/2}$ and $t_0 = 0.57$ μs due to the bandwidth of the measurement circuit [23].

# Supporting Information

# for

# Radio-frequency detected fast charge sensing in undoped silicon quantum dots

Akito Noiri[†,‡,*], Kenta Takeda[†,‡], Jun Yoneda[†,#], Takashi Nakajima[†],
Tetsuo Kodera[¶] and Seigo Tarucha[†,*]

[†]*RIKEN, Center for Emergent Matter Science (CEMS), Wako-shi, Saitama 351-0198, Japan*
[¶]*Department of Electrical and Electronic Engineering, Tokyo Institute of Technology, O-okayama, Meguro-ku, Tokyo 152-8552, Japan*

[#]*Present address: School of Electrical Engineering and Telecommunications, University of New South Wales, Sydney, New South Wales 2052, Australia*

[‡]These authors contributed equally to this work.
[*]e-mail: akito.noiri@riken.jp or tarucha@riken.jp

Supporting Information is provided on single-spin readout by energy selective tunneling and single-spin control.

## 1. Single-spin readout by energy selective tunneling

In the main text, we discuss the performance of singlet-triplet readout. Spin-to-charge conversion by the energy selective tunneling [1,2] is another scheme for single-spin readout widely used in spin qubit experiments and therefore we discuss the performance of this scheme in terms of SNR. Figure S1 shows a typical real time trace of charge sensor signal for spin detection measured near the boundary between (0,1) and (1,1) charge states (shown by M in Figure S2(a)). When the spin state is up, a blip is observed in the time trace associated with electron tunneling events between the left QD and the adjacent reservoir. Normally, the signal of this scheme, the difference of charge sensor signal between (0,1) and (1,1) states, is greater than that of the singlet-triplet readout (Figure 3(c)). In fact, we observe the signal level and therefore the SNR [the noise level is almost unchanged] are enhanced by a factor of 2.5 compared to that of the singlet-triplet readout.

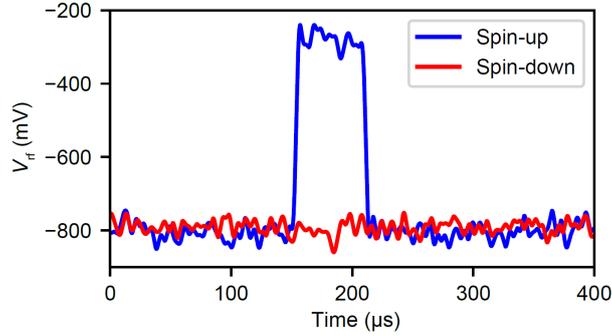

**Figure S1.** Real time traces of charge sensor signal for spin detection. The blue (red) trace shows a typical signal in the case of spin-up (spin-down). The device and measurement setup are the same as those used in Figure 3. The sensor signal is low-pass filtered at a cut-off frequency of 100 kHz.

## 2. Single-spin control

In this section, we demonstrate the compatibility of our small accumulation gate device with a high performance of spin qubit operation. This is done in a different gate voltage condition from that used in Figure 3 with a relatively small inter-dot tunnel coupling, where spin initialization by slow adiabatic passage becomes inefficient [3,4]. In order to initialize and measure the single-spin state, we instead use spin-dependent electron tunneling between the left QD and the adjacent reservoir (see Figures S2(a), (b) and S1). The device has a micro-magnet on top of the accumulation gate for electric dipole spin resonance (EDSR) manipulation of an electron spin [5,6]. The DQD is controllable and stable enough to perform a single qubit rotation experiment. Indeed, we observe a clear Rabi oscillation by applying a microwave (MW) burst to the gate electrode C as shown in Figure S2(c).

From this result, we confirm that the small accumulation gate design is compatible with a high-fidelity spin qubit operation.

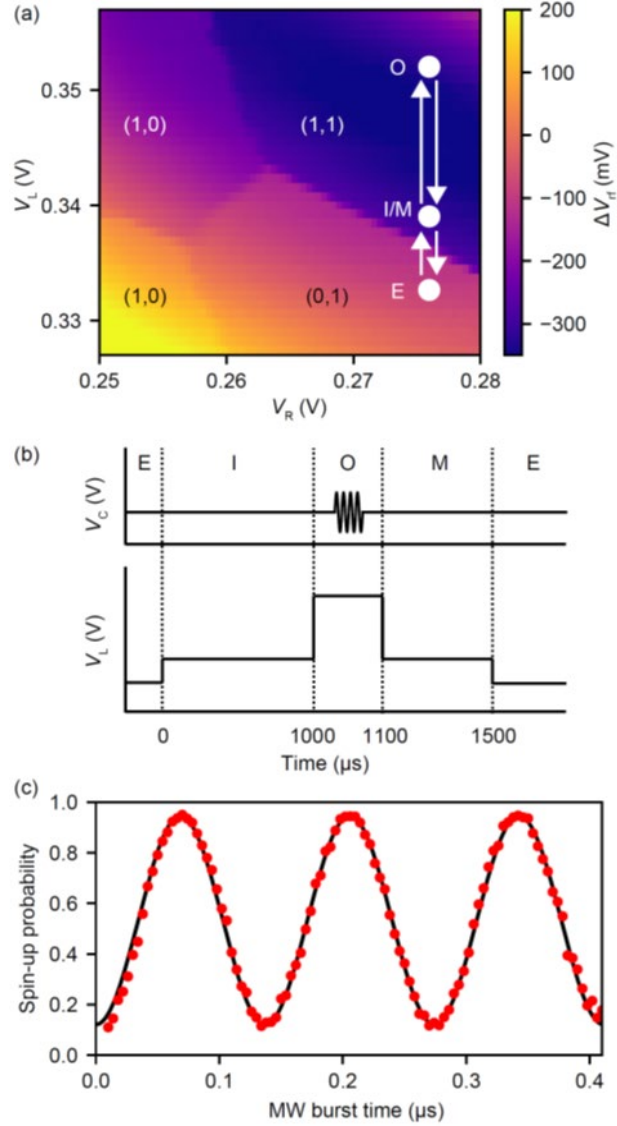

**Figure S2.** Rabi oscillation of a single-spin in the left QD by EDSR. (a) Stability diagram showing gate voltage conditions where evacuation (E), initialization (I), operation (O) and measurement (M) are performed. (b) Schematic of the voltage pulse shape. (c) Rabi oscillation measured at the external magnetic field of 0.54 T and the MW frequency of 16.514 GHz. From a sine curve fit, we obtain a Rabi frequency of 7.3 MHz.